\documentstyle[epsfig]{article}

\newcounter{eqnletter}[equation]
\catcode`\@=11
\@addtoreset{equation}{section}
\catcode`\@=12

\begin{document}
{\centerline {\LARGE {\bf  Real symmetric random matrices and paths counting }}}
\vskip 1 cm
\centerline  {Giovanni M.Cicuta }
\vskip .5 cm
{\small \centerline  { Dip. di Fisica, Univ.di Parma, Parco Area delle Scienze 7A, 43100 Parma, Italy }
\centerline  {and INFN, Sez.di Milano, Gruppo di Parma}
\centerline  {email: cicuta@fis.unipr.it}
 
\vskip .7 cm
{\centerline{\bf Abstract}}
\noindent
Exact evaluation of $<{\rm Tr}\, S^p>$ is here performed for  real symmetric matrices $S$ of arbitrary order $n$, up to some integer $p$, where the matrix entries are independent identically distributed random variables, with an arbitrary probability distribution. 
 These expectations are polynomials in the moments of the matrix entries ; they provide useful information on the spectral density of the ensemble in the large $n$ limit. They also are  a straightforward tool to examine a variety of rescalings of the entries in the large $n$ limit. \\

\section{Introduction}

In the past decades the study and applications of random matrix theory   in a variety of fields of theoretical physics has witnessed a tremendous expansion. In most cases one is interested in universality properties of the model, that is properties of the model, in the limit of infinite order of the matrices, 
which are independent of the detailed form of probability distribution of the random variables. \\ 

Two main classes of models were studied:\\

1) Models with independent entries. In this class of models the (independent) matrix entries are independent identically distributed random variables, that is the joint probability density factorizes.\\
Several matrix ensembles with real or complex entries were studied. From the simplest, where no structure of the matrix is considered, to
the sparse matrices, see for example \cite{seme},
\cite{ak1}, \cite{ak2} ; 
the laplacian matrices (where the diagonal elements are $s_{j,j}=-\sum_{k=1 \to n \, , \, k \neq j}s_{j,k} $), 
see for example \cite{sta}, \cite{for} ;
the band matrices, see for example \cite{wi}, \cite{ak3}, \cite{luca}, \cite{syl}, \cite{pas} ; the block band matrices, see for example \cite{zee1}, \cite{luca2}.\\

2) Unitary-invariant models. The three classic models are real symmetric matrices, complex hermitian matrices or matrices where the entries are real quaternions. Here the joint probability density of the random variables is a function invariant under  unitary (real or complex) transformations. The most studied models  have
the joint probability density in the form of a Boltzmann factor which, for real symmetric matrices, has the form
\begin{eqnarray}
p\left(s_{1,1} \,,\,s_{1,2}\, ,..,s_{n,n}\right) =e^{-{\rm Tr}\,V(S)}
 \label{i.2}
\end{eqnarray}
where $V(x)$ is a polynomial in the variable $x$. The important case where $V(x)=x^2$ describes the gaussian orthogonal ensemble, which is
a model also of independent entries. \\

Further important  matrix ensembles like the Wishart matrices were studied with probability density which often do not belong to the above two classes.\\

Techniques to study the two classes of models are quite different. In the second class, by using the unitarity invariance, the model is rewritten in terms of the eigenvalues of the random matrix. Powerful analytic tools are available, like the technique of orthogonal polynomials. It was found that while the limiting spectral density is sensitive to the details of $V(x)$, several important features related to the  local statistics of eigenvalues exhibit universality properties.\\

In the models of the first class few analytical tools are available and often the limiting spectral density is obtained by numerical evaluation from  samples of large matrices. The most famous result is the universality property of the spectral density  proved by E.Wigner \cite{wi} : with suitable assumptions (which will be reviewed later) the limiting spectral density is the semi-circle, independent from the probability distribution of the entries of the random matrix.\\

In this paper few simple ensembles of real symmetric matrices of the first class will be analyzed : the $\{S \}$ ensemble of real symmetric matrices , the closely related  $\{S_0 \}$ ensemble ( that will be defined in next section), the random sparse matrices and the bi-diagonal matrices ( the simplest type of band matrices).\\

Let $\{S\}$ be the ensemble of real symmetric matrices of order $n$. Any matrix $S \in \{S\}$ is defined by $n(n+1)/2$ real independent matrix elements , $s_{jk}$. The joint probability density factorizes
\begin{eqnarray}
p\left(s_{1,1} \,,\,s_{1,2}\, ,..,s_{n,n}\right) =\prod_{i \leq j}p\left(s_{i,j}\right)
 \label{i.1}
\end{eqnarray}

Let $x$ denote any of the independent entries of the random matrix,   $p(x)$ its probability density,  and $<x^k>$  its moments
\begin{eqnarray}
E\left[x^k\right]=<x^k>=\int dx \, x^k\,p(x)
 \label{i.3}
\end{eqnarray}
I shall only assume that the moments of any degree exist and in the next section 
the expectations E $\left[{\rm Tr} S^p \right]$  are exactly evaluated  as polynomials in   the moments $<x^k>$ for any value of the matrix size $n$ and for $p$ up to $7$ for the ensemble $\{S\}$, up to $p=8$ for the related ensemble $\{S_0\}$, any $p$ for the bi-diagonal matrices.  After a proper n-dependent rescaling, the evaluation of E $\left[{\rm Tr} S^p \right]$, give the moments of the limiting spectral density, if E$(x)=0$.
 Symbolic manipulation software is efficient in sorting out words of given type  and counting them. Then the exact evaluation in the equations (\ref{b.4}),(\ref{b.2}) may be obtained by such software, by using matrices  with symbolic entries of several enough orders, say $n=4, 5, 6, 7, 8, 9$  enough to determine the polynomials in the $n$ variable.
 However computer memory limits a brute force approach, when the number of words is a few hundred thousands. In the next section a simple description is given in terms of (non-markov)  paths which encode all the relevant information. I found useful to combine such manual counting with the symbolic software counting.\\

The evaluation of the moments of the spectral density, by counting classes of paths on the graph associated to the matrix has long history, see for example \cite{wi}, \cite{fu}, \cite{sin}, \cite{ak1}, \cite{ak2}. One may see \cite{past}, \cite{zee} for different methods. Usually the aim was to evaluate the number of classes of paths for large $n$ and large $p$, with $n \gg p$.\\

The finite $p$ exact evaluation provided in the next section will probably be useful for two purposes:\\
a) in several models of sparse matrices or laplacian matrices the limiting spectral density is determined numerically by extrapolating evaluations for several values of $n$. In such cases, the knowledge of the first few moments of the spectral density, provided here, may  add useful information.\\
b) the results (\ref{b.4}),(\ref{b.2}) allow a transparent analysis of the $n$-dependent rescaling preliminary to the $n \to \infty$ limit.  Then it becomes evident why universality of the limiting spectral density is obtained in the case of Wigner scaling and not in the case of sparse matrices scaling. This will be shown in the third section  and the possibility of different scalings will be considered.\\
 
\section{Enumeration of paths}

It is well known that to any square matrix $A=\{a_{ij}\}$, $n \times n$ one may associate a graph with $n$ vertices and a directed edge from vertex $i$ to vertex $j$ , corresponding to the matrix element $a_{ij}$. One can the visualize the evaluation of matrix elements of powers of $A$ as sum of paths on the graph. For example $a_{1,4}a_{4,2}a_{2,3}$ corresponds to the three steps path on the graph from vertex $1$ to vertex $3$ going first to vertex $4$ then to vertex $2$.
Similarly the matrix element $(A^3)_{1,3}=\sum_{j,k=1,..,n} a_{1,j}a_{j,k}a_{k,3}$ corresponds to the sum of  $n^2$  paths on the graph from vertex $1$ to vertex $3$ visiting all possible intermediate vertices $j$, $k$.\\

We begin by considering the ensemble $S_0$ of real symmetric matrices of order $n$, with vanishing entries on the diagonal, $s_{j,j}=0$ for
$j=1,2,..,n$. Because $s_{i,j}=s_{j,i}$ instead of drawing two opposite directed edges between the vertices $i$ and $j$, one draws just one non-directed edge. The graph associated to the generic matrix $S_0$ is then the complete graph with $n$ vertices, shown in the left side of Fig.1\\
\noindent
\epsfig{file=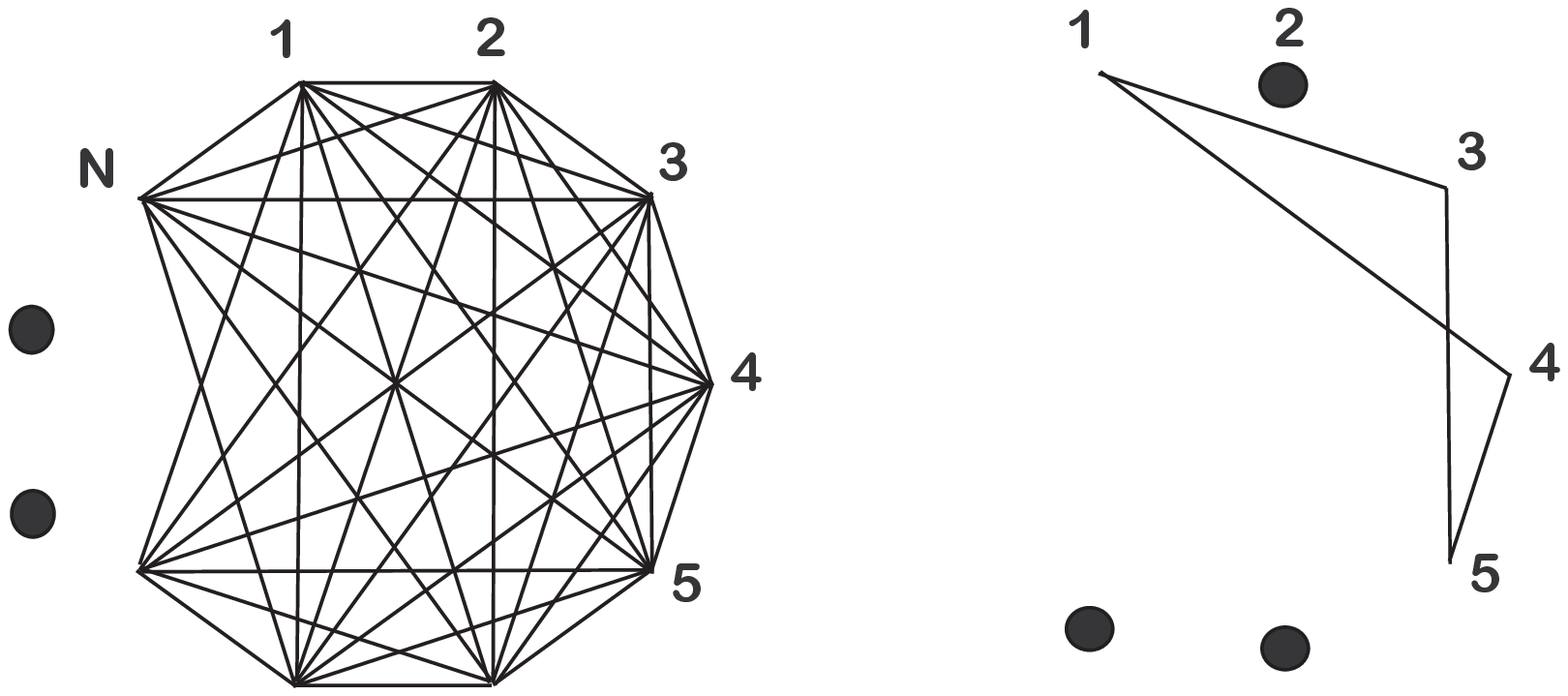,height=5cm}\\
\centerline{Fig.1}\\

The right side of Fig.1 shows the four step path on the graph visiting  the sequence of vertices $\{1,3,5,4,1\}$. It corresponds to the product $s_{1,3}\,s_{3,5}\,s_{5,4}\,s_{4,1}$ which contributes to Tr $S_0^4$. The same path is depicted on the left side of Fig.2, where the horizontal axes represents time, or number of steps.
This type of lattice path is usually called  an excursion or a bridge   
respectively if the path is constrained in the upper half plane or it is not constrained, see for instance \cite{bandirer}. \\

The $n(n-1)/2$ independent entries of a matrix of the ensemble $\{S_0\}$ are independent identically distributed 
  random variables. Let us denote with $<v^k>$ the $k$-moment of the probability density. Then
\begin{eqnarray}
E\left[s_{1,3}\,s_{3,5}\,s_{5,4}\,s_{4,1}\right]=<v>^4 \qquad
 \label{g.1}
\end{eqnarray}
Of course the same contribution is obtained from any other four steps path from vertex one to itself, such that the four entries $s_{i,j}$ are all different. Because of the symmetry of the complete graph, the same is true for any path beginning and ending at a vertex different from one. One is led to a more efficient representation of paths which keeps only the relevant information. \\
\noindent
\epsfig{file=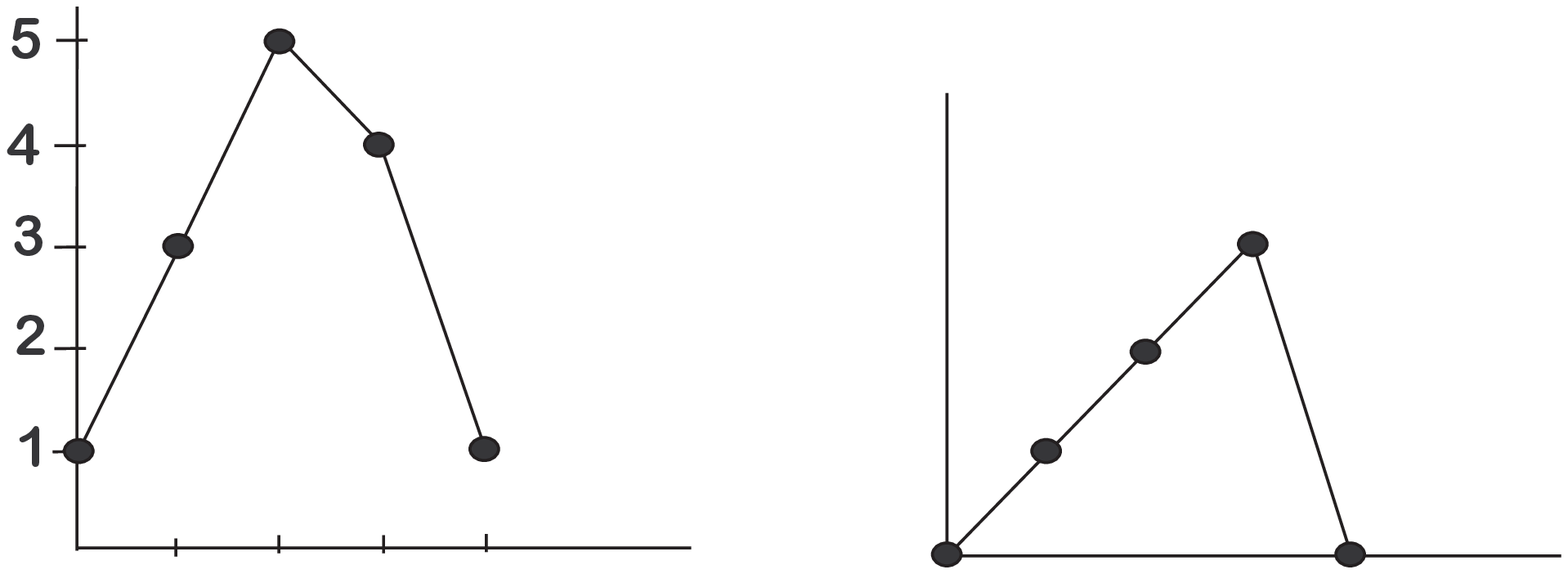,height=4.5cm}\\
\centerline{Fig.2}\\

The paths on the left side of Fig.2 is redrawn on the right side of Fig.2, again with time on the horizontal axis. It reads : start from an arbitrary vertex, next a different vertex, next a vertex different from the two previous ones, next a vertex different from the three previous ones, finally the first vertex. It is trivial to evaluate the number of paths of this kind and obtain the contribution
\begin{eqnarray}
n(n-1)(n-2)(n-3)\,<v>^4
\nonumber
\end{eqnarray}
I shall call reduced paths the graphs, like the right side of Fig.2, which correspond to a multitude of paths where the visited sites are listed, like in the left side of Fig.2.
To evaluate $< {\rm Tr} \,S_0^4>$,
 only four  reduced paths , shown in Fig.3, are needed.   Their contribution is
\begin{eqnarray}
<{\rm Tr}\,S_0^4> &=&n(n-1)\,<v^4>+2 n(n-1)(n-2)\,<v^2>^2+\qquad\nonumber\\
&+& n(n-1)(n-2)(n-3)\,<v>^4
 \label{g.2}
\end{eqnarray}
\noindent
\epsfig{file=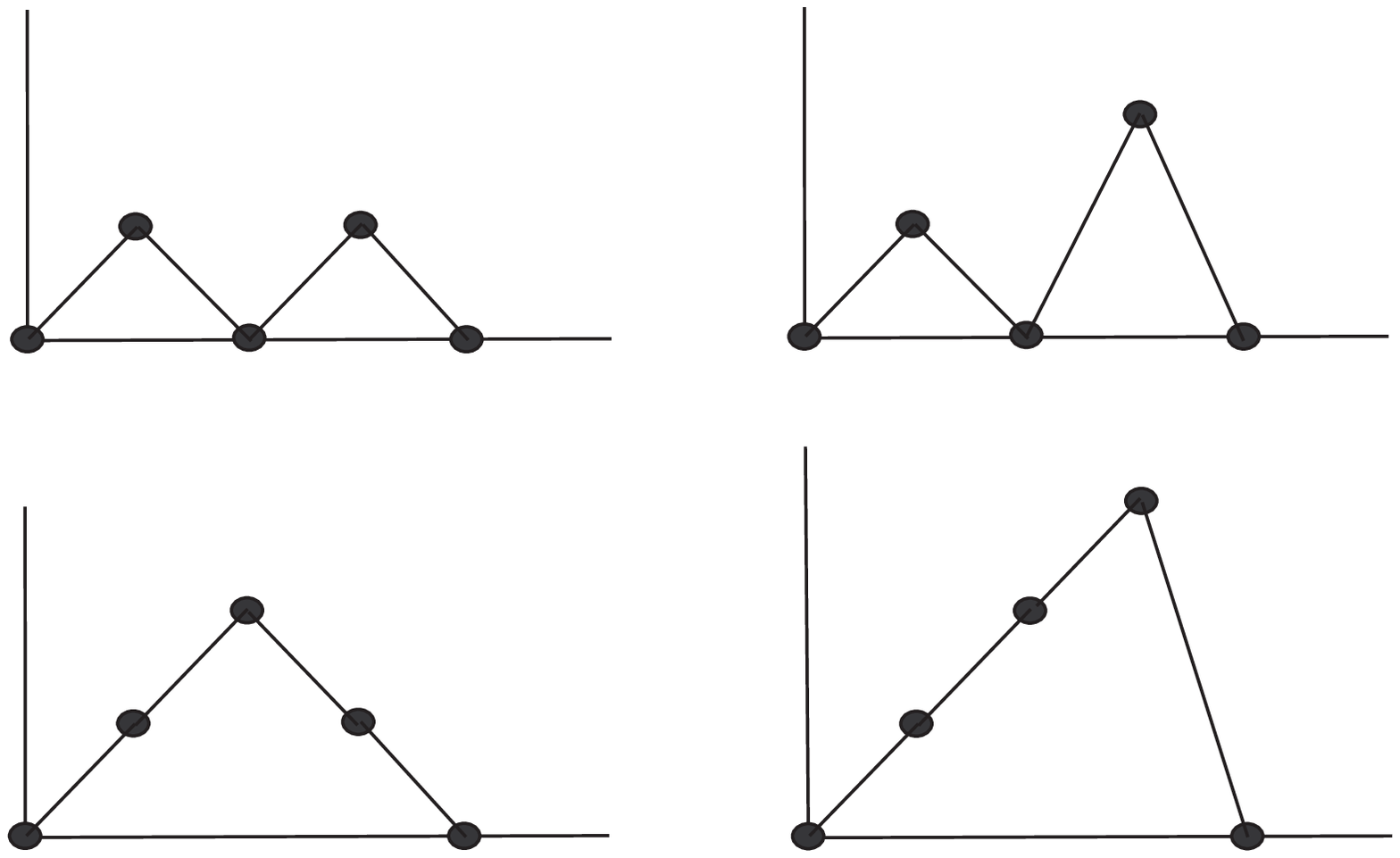,height=7cm}\\
\centerline{Fig.3}\\

From now on I shall only refer to these type of paths.

I shall call  closed paths those where the final vertex is the same as the beginning vertex. Open paths have a final vertex different from the beginning vertex. Only closed paths contribute to the evaluation of Tr $S_0^p$, but open paths are equally important because all the closed paths with $p$ steps are obtained by considering all the open paths with $p-1$ steps and adding one further edge connecting the last vertex to the first vertex (closing the path).\\

Let $S_c(n,q)$ denote the number of closed paths with $n$ steps and such that the maximal altitude reached during the path is $q$. It means that the path visited a total of $q$ distinct vertices of the complete graph, including the initial vertex. Similarly $S_o(n,q)$ will denote the number of open paths with $n$ steps and such that the maximal altitude reached during the path is $q$. The previous remark translates into

\begin{eqnarray}
S_c(n,q)=S_o(n-1,q) \qquad \qquad
 \label{g.1aa}
\end{eqnarray}

One may evaluate the total number of paths , that is the open plus closed paths for any number of steps.\\
 
Consider a path with $n-1$ steps and denote by $y_k$ , $k=0,1,\cdots,n-1$, the altitude of the path at time $k$. Let $q$= maximum$\{y_k\}$. One further step may be added to the path, by connecting $y_{n-1}$ with $y_n$, where $y_n $  is any of the $q+1$ integers $0, 1, 2,..,q+1$, excluding $y_n= y_{n-1}$. Of course there is no path where $n<q$.\\

Let $S(n,q)=S_c(n,q)+S_o(n,q)$ be the total number of paths with $n$ steps and maximal altitude $q$. The previous construction of paths of $n$ steps from paths with $n-1$ steps , translates into the equations

\begin{eqnarray}
&&S(n,q)= q\,S(n-1,q)+S(n-1,q-1) \qquad , \qquad \qquad \nonumber\\
&&S(n,q)=0 \qquad {\rm if} \qquad n<q \qquad ; \qquad S(n,1)=1 \qquad  
\qquad
 \label{g.1a}
\end{eqnarray}

The positive integer numbers $S(n,q)$ are known as Stirling number of the second kind. Their sum over the values of $q$ are known as the Bell numbers $B(n)$, see for example \cite{st}, \cite{an}. Useful formulas are
\begin{eqnarray}
S(n,q)&=&\frac{1}{q!}\sum_{r=1}^q (-1)^{q-r}\left( q \atop r \right) r^q \qquad , \qquad \nonumber\\
B(n)&=&\sum_{q=1}^n S(n,q) \qquad , \qquad \nonumber\\
B(n+1) &=& \sum_{r=0}^n \left( n \atop r \right) B(r) \qquad , \qquad \nonumber\\
{\rm exp}\left(e^x-1 \right)&=&\sum_{n=0}^{\infty}\frac{1}{n!}\, x^n \,B(n) \qquad \qquad
 \label{g.2a}
\end{eqnarray}
For example the total number of paths with four steps is $B(4)=15$ and $S(4,1)=1$, $S(4,2)=7$, $S(4,3)=6$, $S(4,4)=1$. \\
The first few Bell numbers are $1\, ,\,2\, ,\,5\,,\,15\,,\,52\,,\,203\,,\,877\,,\,4140\,,\cdots$\\
Because of eq.(\ref{g.1aa}), $S_c(n)=S_o(n-1)=B(n-1)-S_c(n-1)$, then the first few numbers of closed paths are 
\begin{eqnarray}
 &&S_c(2)=1 \quad , \quad S_c(3)=1  \quad , \quad S_c(4)=4
 \quad , \quad S_c(5)=11  \quad , \quad S_c(6)= 41 \qquad , \nonumber\\
&&S_c(7)=162 \quad , \quad S_c(8)=715 \quad , \cdots \qquad
 \label{g.2aa}
\end{eqnarray}

By substituting all the non-vanishing entries of the matrix $S_0$ with one, a circulant matrix is obtained where it is easy to evaluate
\begin{eqnarray}
{\rm Tr}\, S_0^{2p}&=&(n-1)[(n-1)^{2p-1}+1] \qquad , \nonumber\\
{\rm Tr}\, S_0^{2p+1}&=&(n-1)[(n-1)^{2p}-1]
 \label{g.6}
\end{eqnarray}
One obtains a new and easy determination of the number of closed paths $S_c(n,q)$ by comparing the above numbers with the contribution  of the graphs. For instance, in the case of paths with $8$ steps, by solving the equation
\begin{eqnarray}
(n-1)\left( (n-1)^7+1\right)=\sum_{q=1}^8 n(n-1)(n-2)\cdots (n-q)\,S_c(8,q)
\qquad
\label{g.7}
\end{eqnarray}
We obtain
\begin{eqnarray}
S_c(8,1)=1 \qquad  &,& \qquad S_c(8,2)=42 \qquad , \qquad S_c(8,3)=231 \qquad , \qquad \nonumber\\
S_c(8,4)=294 \qquad &,& \qquad S_c(8,5)=126 \qquad , \qquad  S_c(8,6)=20 \qquad , \qquad \nonumber\\
S_c(8,7)=1 \qquad &,& \qquad \sum_{q=1}^8S_c(8,q)=715 \qquad \qquad
\label{g.8}
\end{eqnarray}

Finally one draws the graphs and reads their contributions. It is a time consuming activity, helped  by symbolic manipulation software. I report here the evaluations up to E $[{\rm Tr} \,S_0^8]$, which results from $715$ graphs.\\ 

\begin{eqnarray}
\frac{1}{n} <{\rm Tr} \,S_0>&=&\,0\qquad , \qquad \nonumber\\
\frac{1}{n} <{\rm Tr} \,S_0^2>&=&\,(n-1)<v^2>\qquad , \qquad \nonumber\\
\frac{1}{n} <{\rm Tr} \,S_0^3>&=&\,(n-1)(n-2)<v>^3 \qquad , \qquad 
\nonumber\\
\frac{1}{n} <{\rm Tr} \,S_0^4>&=&\,(n-1)<v^4>+2(n-1)(n-2)<v^2>^2+\qquad  \qquad  \nonumber\\
&+&(n-1)(n-2)(n-3)<v>^4\qquad , \qquad 
\nonumber\\
\frac{1}{n} <{\rm Tr} \,S_0^5>&=&\,5(n-1)(n-2)<v^3><v>^2+ \qquad  \qquad \nonumber\\
&+& 5(n-1)
(n-2)(n-3)<v^2><v>^3+\qquad , \qquad \nonumber\\
&+&(n-1)(n-2)(n-3)(n-4)\,<v>^5\qquad ,  \nonumber\\
\frac{1}{n} <{\rm Tr} \,S_0^6>&=&\,(n-1)<v^6>+6(n-1)(n-2)<v^4><v^2>+
\nonumber\\
&+&(n-1)(n-2)(5n-11)\,<v^2>^3  +\qquad \nonumber\\
&+&6(n-1)(n-2)(n-3)\, <v^3><v>^3 +\qquad \nonumber\\
&+&3(n-1)(n-2)(n-3)(2n-5)\,<v^2><v>^4 +\qquad \nonumber\\
&+&(n-1)(n-2)^2(n-3)(n-4)\, <v>^6 \qquad ,\nonumber\\
\frac{1}{n} <{\rm Tr} \,S_0^7>&=&\,7(n-1)(n-2)\,<v^5><v>^2+
14(n-1)(n-2)\,<v^3>^2<v>+\qquad \nonumber\\
&+& 7(n-1)(n-2)(n-3)\,<v^4><v>^3+\qquad \nonumber\\
&+& 35(n-1)(n-2)(n-3)\,<v^3><v^2><v>^2+ \qquad \nonumber\\
&+& 7(n-1)(n-2)(n-3)^2(3n-8)\,<v^2>^2<v>^3+\qquad \nonumber\\
&+& 7(n-1)(n-2)(n-3)(n-4)\,<v^3><v>^4 +
\qquad \nonumber\\
&+& 7(n-1)(n-2)^2(n-3)(n-4)\,<v^2><v>^5+
\qquad \nonumber\\
&+& (n-1)(n-2)(n-3)(n-4)(n^2-4n+2)\,<v>^7 \qquad ,
\qquad \nonumber\\
\frac{1}{n} <{\rm Tr} \,S_0^8>&=&\,(n-1)<v^8>+(n-1)(n-2)[8<v^6><v^2>+6<v^4>^2]+
\qquad \nonumber\\
&+& (n-1)(n-2)^2 28 <v^4><v^2>^2+\qquad \nonumber\\
&+& (n-1)(n-2)(n-3)[8<v^5><v>^3+20<v^3>^2<v>^2]+
\qquad \nonumber\\
&+&(n-1)(n-2)^2(n-3)48<v^3><v^2><v>^3+
\qquad \nonumber\\
&+& 4(n-1)(n-2)(n-3)(2n-3)<v^4><v>^4+
\qquad \nonumber\\
&+& (n-1)(n-2)(n-3)(14n-19) <v^2>^4+
\qquad \nonumber\\
&+& 2(n-1)(n-2)(n-3)(14n^2-66n+51)<v^2>^2<v>^4+
\qquad \nonumber\\
&+& 8 n(n-1)(n-2)(n-3)(n-4)<v^3><v>^5+
\qquad \nonumber\\
&+& 8(n-1)(n-2)(n-3)(n-4)(n^2-3n-2)<v^2><v>^6+
\qquad \nonumber\\
&+& (n-1)(n-2)(n-3)(n-4)(n-5)(n^2-n-4)<v>^8 \qquad
\label{b.4}
\end{eqnarray}

M.Bauer and O.Golinelli, in their analysis of the incidence matrix for random graphs \cite{bau},  evaluated traces of powers of a random matrix by evaluating normalized $k$-plets. These are sequences of $k$ positive integers $(v_1, v_2,\cdots , v_k)$ such that :\\
1) $v_1 \neq v_2$ , $v_2 \neq v_3$, $\cdots$,$v_{k-1} \neq v_1$\\
2) if $v_{\beta}>1$ is in a normalized $k$-plets, then also $v_{\beta'}=v_{\beta}-1$ is in the same normalized $k$-plet and  $\beta'<\beta$.\\    
The normalized $k$-plets are in one-to-one correspondence with the reduced closed paths described in this paper : the sequence of the $k$ integers of a normalized $k$-plet is the sequence of the  $y$-coordinates 
$\{y_0=1, y_2,\cdots , y_{k-1} \}$, of a reduced closed paths with $k$ steps. 
For example, the four reduced paths in Fig.3 correspond to the four normalized $4-$plets $\{1,2,1,2\}$ , $\{1,2,1,3\}$ , $\{1,2,3,1\}$ , $\{1,2,3,4\}$. 
I prefer the graph representation because I find it easier to read the repeated steps and it allows several generalizations  for random matrices with diagonal elements not equal to zero, which may have the same probability distribution of the other entries, as in the next section, or a different probability distribution, as it occurs in laplacian matrices, not described in this paper.\\

 \vskip 1cm

{\bf From the ensemble $\{S_0\}$ to the ensemble $\{S\}$}.\\

The graph corresponding to a matrix of the ensemble $\{S\}$}, of order $n$, is the complete graph with $n$ vertices, with a loop added to each vertex. It is depicted  in Fig.4.\\
\noindent
\epsfig{file=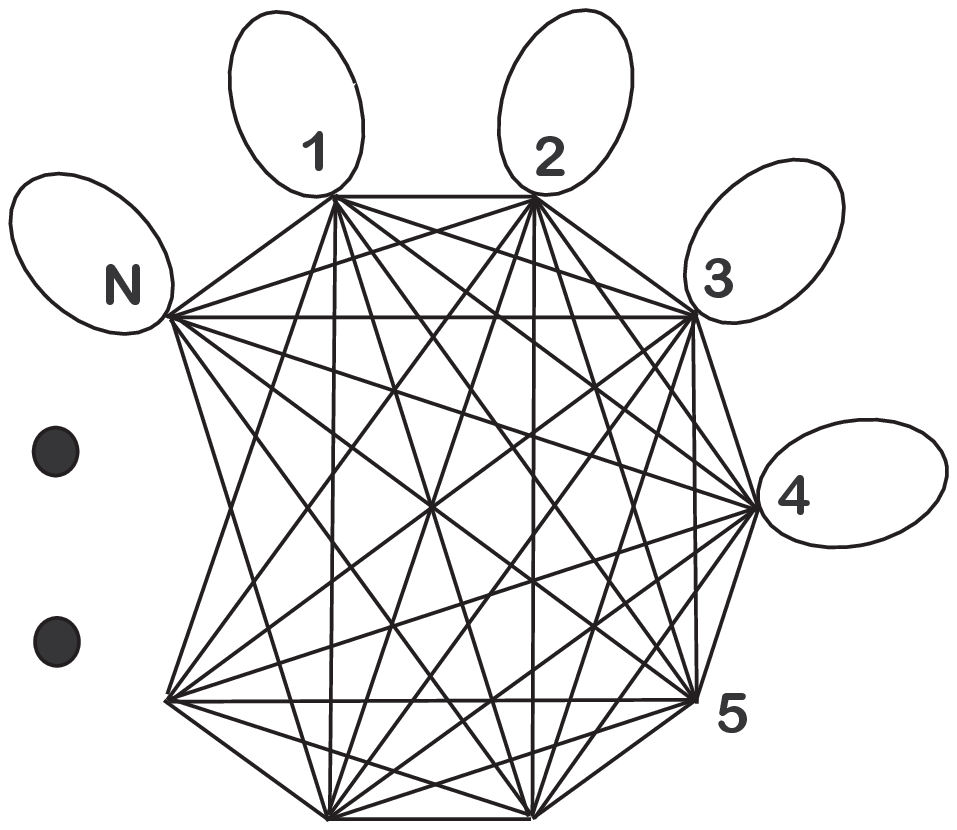,height=5cm}\qquad
 Fig.4\\

The evaluation of E $\left[{\rm Tr}\,S^p \right]$ 
may be performed by considering  the closed graphs with $p$ steps where any number of horizontal steps are now allowed at any vertex. 
A simple example will show how the new reduced paths may be obtained from
the reduced paths contributing to the evaluation of E $\left[{\rm Tr}\,S_0^k \right]$ , $k=0,1,\cdots ,p$ , (where horizontal steps are missing) by adding the $p-k$ horizontal steps in all possible ways.\\
For example to evaluate  E $\left[{\rm Tr}\,S^4 \right]$
we  consider the two paths, in Fig.5, respectively of $2$ and $3$ steps.\\
\noindent
\epsfig{file=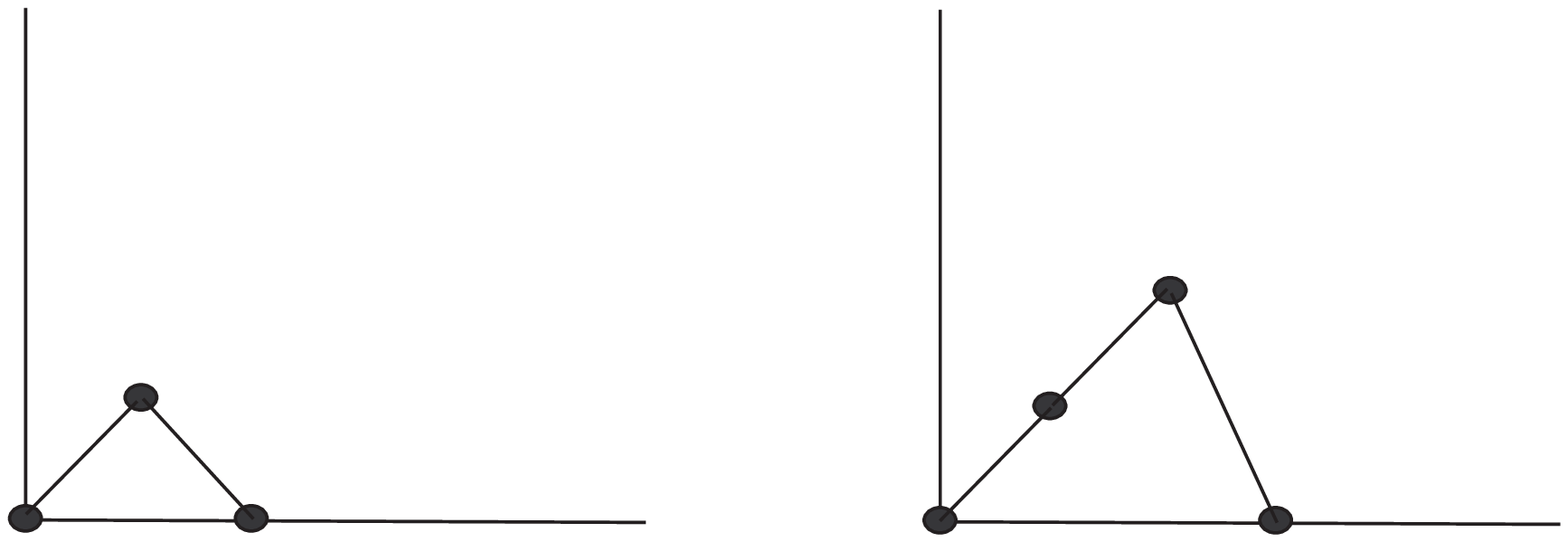,height=3cm}\qquad  Fig.5\\

Let us first consider the graph of $3$ steps. It has $4$ vertices and we  obtain $4$ closed graphs contributing to E $\left[{\rm Tr}\,S^4 \right]$ by
adding one horizontal step  in four possible ways , as it is shown in Fig.6.  Each of the $4$ graphs corresponds to the contribution
\begin{eqnarray}
n(n-1)(n-2)\,<v>^4
\nonumber
\end{eqnarray}
\noindent
\epsfig{file=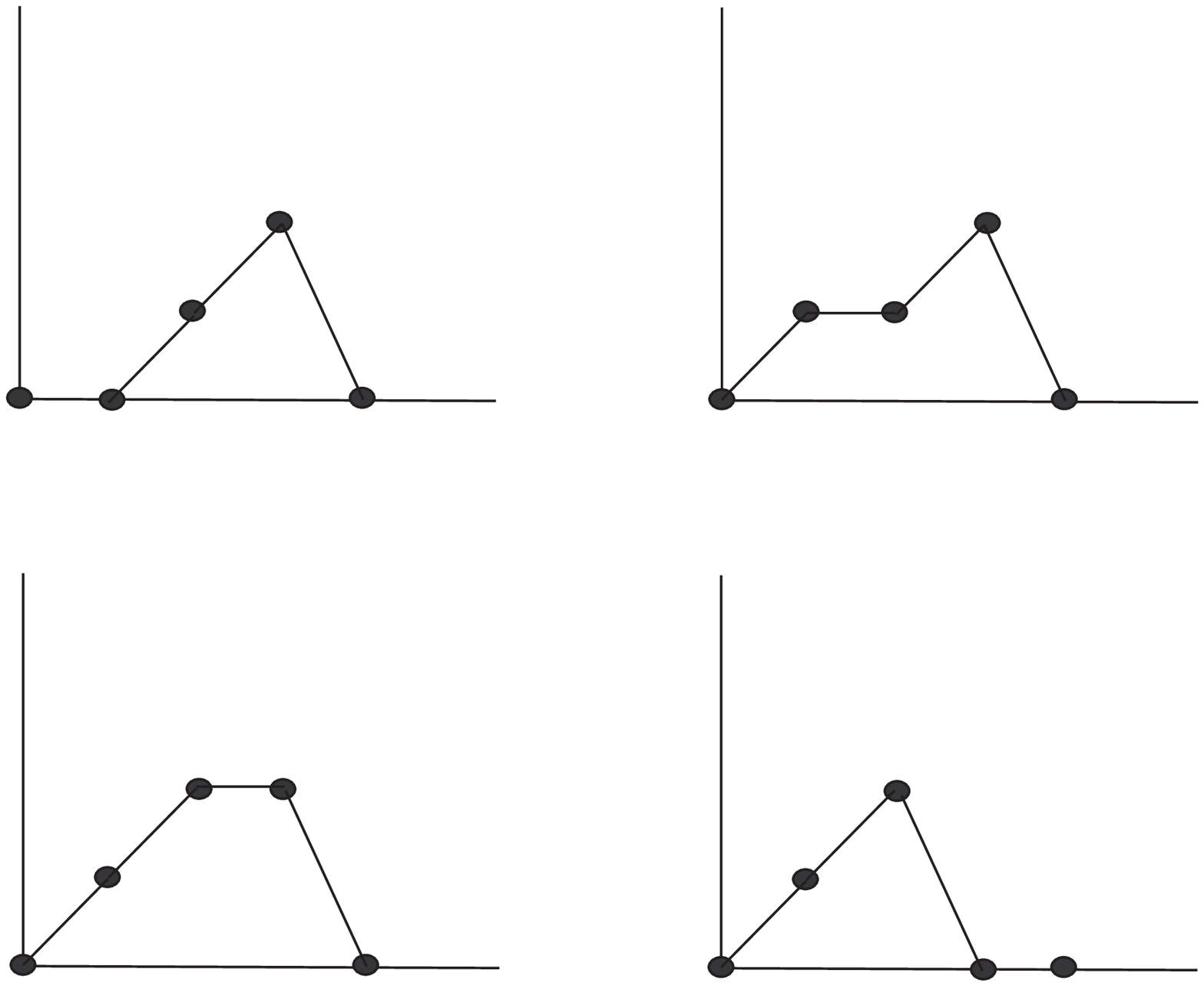,height=5.5cm}\qquad Fig.6\\

Next one adds two horizontal steps to the graph of $2$ steps in Fig.5, in $6$ possible ways, thus obtaing the $6$ graphs shown in Fig.7. In the first $4$ of them , the two horizontal steps are at the same vertex and their contribution is $4n(n-1)<v^2>^2$. The last $2$ graphs have   the two horizontal steps in different vertices, then their contribution is $2n(n-1)<v^2><v>^2$ \\

\epsfig{file=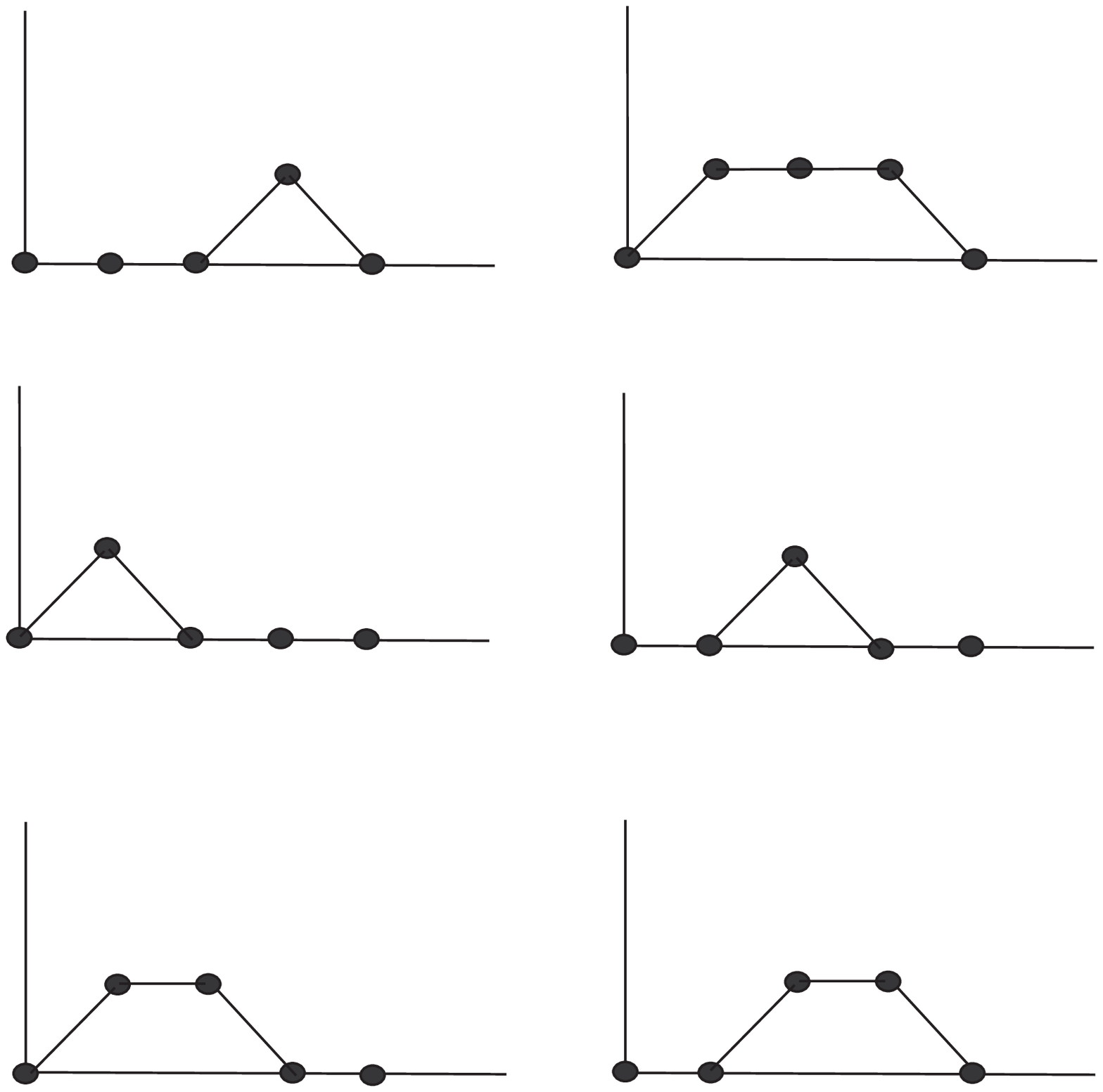,height=7cm}\qquad Fig. 7\\
Finally one adds the contribution of one graph with four  horizontal steps , that is
$n<v^4>$. By collecting the new terms one obtains
 \begin{eqnarray}
 E \left[{\rm Tr}\,S^4 \right]&=& n<v^4>+4n(n-1)<v^2>^2+2n(n-1)<v^2><v>^2+ \qquad \nonumber\\
&+&
4n(n-1)(n-2)<v>^4+ E \left[{\rm Tr}\,S_0^4 \right]=\qquad \nonumber\\
&=&n^2\,<v^4>+
2n(n-1)<v^2><v>^2+
2n^2(n-1)<v^2>^2+\qquad \nonumber\\
&+&n(n^2-1)(n-2)\,<v>^4
\label{b.5}
\end{eqnarray}

\noindent
As a trivial check, by replacing all the moments $<v^k> \to 1$ , one finds  E$\left[{\rm Tr}\,S^4 \right]=n^4$. \\
One may then evaluate
 
\begin{eqnarray}
\frac{1}{n} <{\rm Tr} \,S>&=&<v>\qquad , \qquad \nonumber\\
\frac{1}{n} <{\rm Tr}\, S^2>&=& n\,<v^2> \qquad , \qquad \nonumber\\
 \frac{1}{n} <{\rm Tr} \,S^3>&=&<v^3>+3(n-1)<v^2><v>+(n-1)(n-2)<v>^3 \qquad , \qquad 
\nonumber\\
\frac{1}{n} <{\rm Tr} \,S^4> &=& n\,<v^4>+ 2(n-1)<v^2><v>^2+\nonumber\\&+& 2n(n-1)<v^2>^2+\nonumber\\ 
&+&(n^2-1)(n-2)<v>^4 \qquad , \qquad \nonumber\\
\frac{1}{n} <{\rm Tr}\, S^5> &=&<v^5>+5(n-1)<v^4><v>+  5(n-1)<v^3><v^2>+\nonumber\\
&+&5(n-1)(2n-3) <v^2>^2<v>+5(n-1)(n-2)<v^3><v>^2+ \nonumber\\
&+&
5(n-1)(n-2)^2<v^2><v>^3+ \nonumber\\
&+&(n-1)(n-2)(n^2-2n+2)<v>^5\quad ,\nonumber\\
\frac{1}{n} <{\rm Tr}\, S^6> &=& n\,<v^6>+
6(n-1)<v^4><v>^2+6(n-1)<v^3><v^2><v>+\nonumber\\
&+&3(n-1)(2n+1)<v^4><v^2>+\nonumber\\
&+&15(n-1)(n-2)<v^2>^2<v>^2+\nonumber\\
&+&(n-1)(5n^2-6n-5)<v^2>^3+\nonumber\\
&+&6(n-1)(n-2)(n+3)<v^3><v>^3+\nonumber\\
&+&3(n-1)(n-2)(2n^2+n-17) \,<v^2><v>^4+\nonumber\\
&+&(n-1)(n-2)(n^3-3n^2-7n+23)\,<v>^6
\qquad , \qquad \nonumber\\
\frac{1}{n} <{\rm Tr}\, S^7> &=&<v^7>+7(n-1)<v^6><v>+14(n-1)<v^4><v^3>+ \nonumber\\
&+&7(n-1)<v^5><v^2>+\nonumber\\
&+&14(n-1)(3n-4)<v^4><v^2><v>+\nonumber\\
&+& 7(n-1)(n-2) <v^5><v>^2+\nonumber\\
&+& 7(n-1)(3n-5) <v^3><v^2>^2 +\nonumber\\
&+& 14(n-1)(n-2) <v^3>^2 <v> +\nonumber\\
&+& 7(n-1)(n-2)^2 <v^4><v>^3 + \nonumber\\
&+& 7(n-1)(n-2)(5n-8) <v^3><v^2><v>^2 +\nonumber\\
&+& 35(n-1)^2(n-2) <v^2>^3 <v> +\nonumber\\
&+& 7(n-1)(n-2)(n^2+1) <v^3><v>^4+\nonumber\\
&+& 7(n-1)^2(n-2)(3n-8) <v^2>^2<v>^3 +\nonumber\\
&+& 7(n-1)(n-2)(n^3-2n^2-2n-2) <v^2><v>^5 +\nonumber\\
&+& (n-1)(n-2)(n-3)(n^3-n^2-10n-1)<v>^7 \qquad 
\nonumber\\
\label{b.2}
\end{eqnarray}

{\bf The bi-diagonal symmetric matrices.}\\

In a lattice in one dimension with $n$ sites and periodic boundary conditions, each site has two next neighbours, the lattice has $n$  edges. To each edge $(k,k+1)$ we associate the random variable $a_{k,k+1}=a_{k+1,k}$.\\

We consider the real symmetric bi-diagonal random matrix $A$
\begin{eqnarray}
A=\left( \begin{array}{ccccccc}
0& a_{1,2} & 0 & 0 & \cdots & 0 & a_{1,n} \\
a_{1,2} & 0 & a_{2,3} & 0 &\cdots & 0 & 0 \\
0 & a_{2,3} & 0 & a_{3,4} & \cdots & 0 & 0 \\
\cdots & \cdots & \cdots & \cdots & \cdots & \cdots & \cdots \\
0 & 0 & 0 & 0 & a_{1,n-2}   & 0 & a_{n-1,n}  \\
a_{1,n} & 0 & 0 & 0 &\cdots & a_{n-1,n}   & 0 
\end{array} \right) \qquad , \qquad 
\label{c.1}
\end{eqnarray}
A walk on the lattice with $p$ steps may return to the initial site only if $p$ is even. The  number of such paths \footnote{Because of the periodic boundary conditions, there exist walks on the lattice which return to the initial site from the opposite side of the first step,
if $p\geq n$. Since we are interested in the limit $n \to \infty$ with fixed $p$, those walks are here neglected.}
is  $\left( p \atop p/2 \right)$. This is the number of terms   contributing to 
$\frac{1}{n} <{\rm Tr} \,A^p>$.\\
The left side of Fig.8 shows the periodic one dimensional lattice with $n$ vertices, associated to the matrix $A$ in eq.(\ref{c.1}). The right side of Fig.8 shows a path from site $j$ returning to site $j$ after $14$ steps, then contributing to Tr $A^{14}$.\\
Any path may be seen as composed of paths staying (strictly or weakly) in the positive or in the negative half-plane. Paths which do not enter the negative half-plane (these are the weakly positive paths), such as the one depicted in the left side of Fig.9 , are known as Dyck paths. Dyck paths have a one-to-one correspondence with rooted trees, as indicated in the right side of Fig.9. It is the possible to enumerate the number of steps (up or down) between level $k$ and level $k+1$ for the class od Dyck paths of any length. Next by composing weakly positive and weakly negative Dyck paths, one obtains enumeration of steps for the generic paths, see for instance \cite{cic}.

\noindent
\epsfig{file=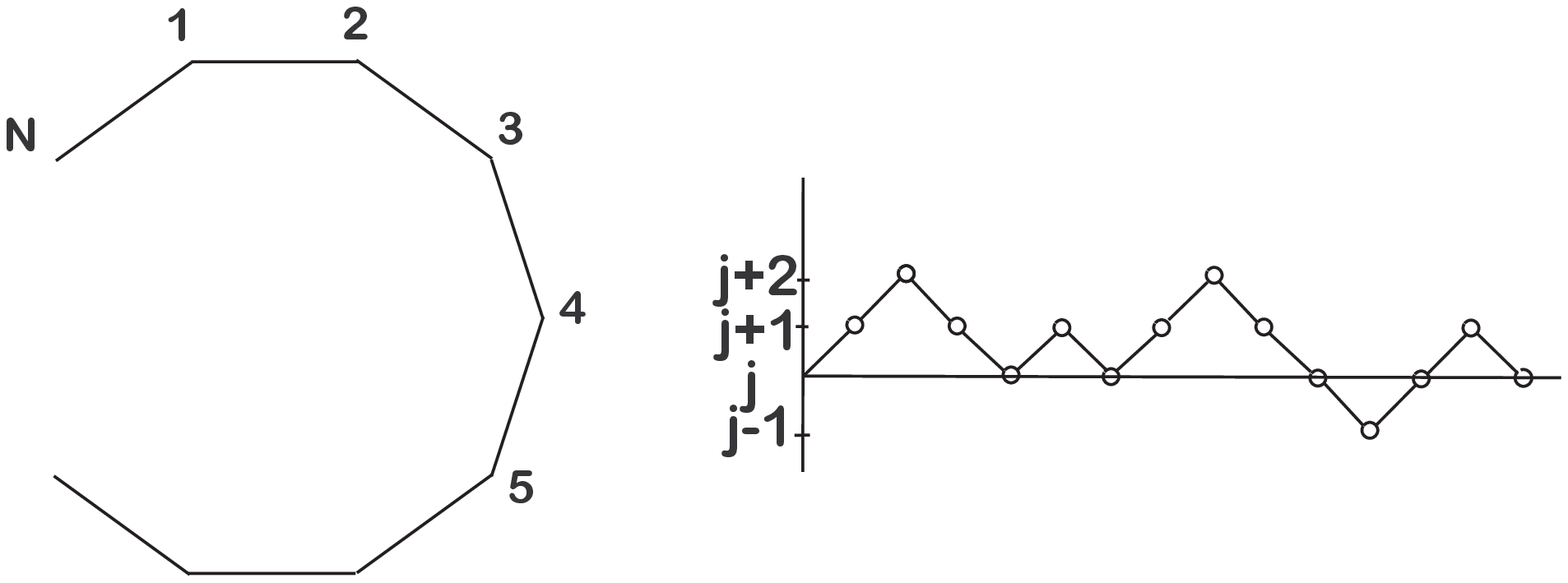,height=4.5cm}\\
\centerline{Fig.8}\\

One easily evaluates
\begin{eqnarray}
\frac{1}{n} <{\rm Tr} \,A^{2p+1}>&=&\,0\qquad , \qquad 
\frac{1}{n} <{\rm Tr} \,A^0>=1\,\nonumber\\
\frac{1}{n} <{\rm Tr} \,A^2>&=&\,2<v^2>\qquad , \qquad \nonumber\\
\frac{1}{n} <{\rm Tr} \,A^4>&=&\,2<v^4>+4<v^2>^2
\qquad , \qquad \nonumber\\
\frac{1}{n} <{\rm Tr} \,A^6>&=&\,2<v^6>+12<v^4><v^2>+6<v^2>^3\qquad , \qquad \nonumber\\
\frac{1}{n} <{\rm Tr} \,A^8>&=&\,2<v^8>+16<v^6><v^2>+12<v^4>^2+\nonumber\\
&+&32<v^4><v^2>^2+8<v^2>^4\qquad , \qquad \nonumber\\
\frac{1}{n} <{\rm Tr} \,A^{10}>&=&\,2<v^{10}>+20<v^8><v^2>+40<v^6><v^4>+\qquad \nonumber\\
&+&50<v^6><v^2>^2+70<v^4>^2<v^2>+60<v^4><v^2>^3+
\qquad \nonumber\\
&+&10<v^2>^5 \qquad \qquad
\label{c.2}
\end{eqnarray}

\noindent
\epsfig{file=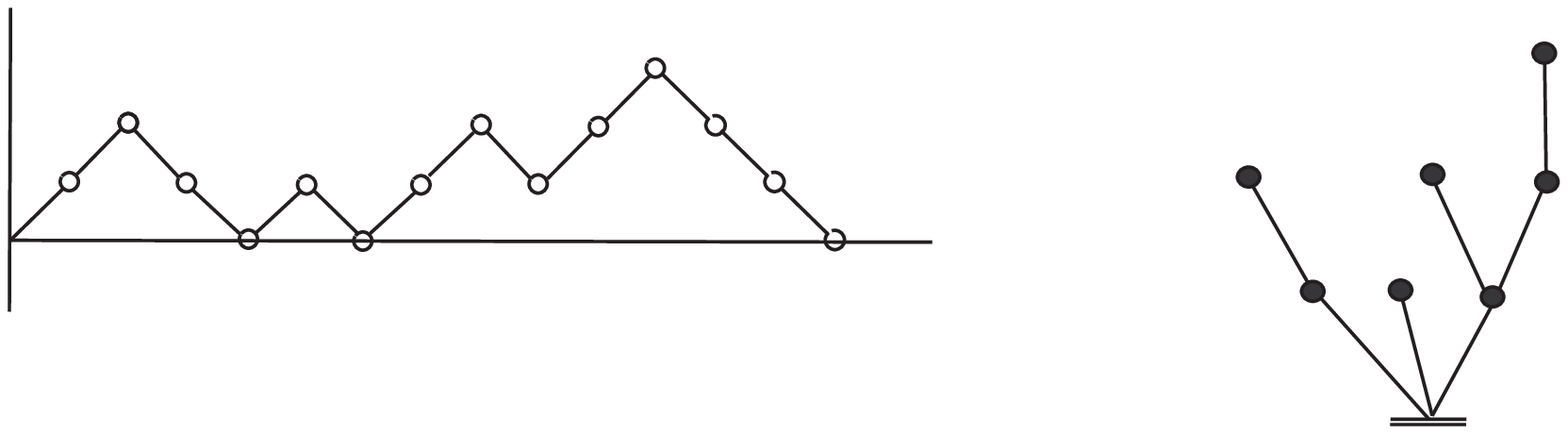,height=4.5cm}\\
 \centerline{Fig.9}\\

Fig.9 shows the one-to-one correspondence between Dyck paths and rooted trees : given the tree, one may follow its external contour, starting at the left side of the root. As one moves up along the first branch, one generates two up steps of the Dyck path , then moving down along the same branch generates the two steps down ; next branch is shorter corresponding to one step up and one step down for the Dyck path. Finally, after touring next branch, one reaches
 the right side of the root.\\

\section{Scalings and/or Universality}
To study  the existence of a limiting spectral density in the $n \to \infty$ limit, 
some $n$-dependent scaling for the entries of the matrices is needed.\\
One may recall that the existence of finite large-$n$ limit of a finite number of rescaled traces of powers of the random matrices is neither necessary nor sufficient condition for the existence of a limiting spectral density. Indeed in the case of gaussian matrices when $<v>$ is not equal to zero and Wigner scaling (which will be recalled here below) the large-$n$ spectral density of the ensemble contains the classical semicircle separated by a small peak, which disappears at $n=\infty$, yet produces the large-$n$ behaviour of the traces.\footnote{I suppose this is well known to experts, but I have not found a proper reference, then I offer the short discussion in the Appendix, by using the addition law for fixed plus random ensembles.}. A more complex distribution with possibly several peaks  occurs with the sparse matrices \cite{bau}, \cite{fu}, \cite{for}.\\

Conversely, a smooth limiting spectral density for the matrix ensemble may exist , with unlimited support, and decreasing as an inverse power for large eigenvalues. Then only a finite number of moments of the spectral density will exist.\\

Then the aim of this section   is not to derive the
limiting spectral density from the 
 table of values (\ref{b.4}), (\ref{b.2}) 
for $< {\rm Tr}\, S^p>$ and $<{\rm Tr} \,S_0^p>$  after proper rescaling, but merely to review the effects of two well known
scalings and to indicate the existence of intermediate scalings which appear to yield the same results of the Wigner scaling for every finite $p$. \\

{\bf Wigner scaling}\\
 The Wigner scaling consists of two assumptions :\\
a) the i.i.d. entries have vanishing expectation value, that is E $(s_{i,j})=<v>=0$ for any $i,j=1,2,\cdots,n$.\\
b)   the i.i.d. entries  are written
 \begin{eqnarray}
&&s_{i,j}=\frac{\xi_{i,j} }{\sqrt{n}} \qquad , \qquad
\xi_{i,j} \quad {\rm independent} \quad {\rm on} \quad n \nonumber\\
&& {\rm then}\qquad {\rm E}\,\left[ (s_{i,j})^k \right]=<v^k>=\frac{1}{n^{k/2}}c_k \qquad , \qquad c_k
\quad {\rm independent} \quad {\rm on} \quad n \qquad \qquad \qquad
\label{cc.1}
\end{eqnarray}
By inserting these assumptions in the tables (\ref{b.4}),
(\ref{b.2}) and by keeping only the leading order contribution in the limit $n \to \infty$ , both tables collapse into the very simple result

 \begin{eqnarray}
\frac{1}{n} <{\rm Tr}\, S^2>&=& \,<\xi^2> \qquad , \qquad \nonumber\\
\frac{1}{n} <{\rm Tr} \,S^4> &=& 2\,<\xi^2>^2+O\left(\frac{1}{n}\right) \qquad , \qquad \nonumber\\
\frac{1}{n} <{\rm Tr}\, S^6> &=&5\,<\xi^2>^3+O\left(\frac{1}{n}\right) \qquad , \qquad \nonumber\\
\frac{1}{n} <{\rm Tr}\, S^8> &=&14\,<\xi^2>^4+O\left(\frac{1}{n}\right) \qquad , \qquad \nonumber\\
\label{cc.2}
\end{eqnarray}
Which are easily seen as the lowest moments of the semi-circle density
 \begin{eqnarray}
\frac{1}{n} <{\rm Tr}\, S^{2k}> &=&C_k\,<\xi^2>^k+O\left(\frac{1}{n}
\right) \qquad , \qquad \nonumber \\
C_k=\frac{1}{k+1}\left( 2k \atop k \right)&=&\frac{1}{2\pi}\int_{-2}^2 dx\, \sqrt{4-x^2}\,x^{2k}
\label{bb.4}
\end{eqnarray}
The variance $\sigma^2=<\xi^2>$ of the probability density of the $\xi_{i,j}$ random variables is the only parameter entering in the limiting spectral density
  \begin{eqnarray}
G(z)&=& \lim_{n \to \infty}\frac{1}{n} {\rm E }\left[{\rm Tr}\frac{1}{z-S}\right] =
\lim_{n \to \infty} \frac{1}{n}\sum_{p=0}^{\infty}\frac{1}{z^{p+1}}{\rm E }\left[{\rm Tr} \,S^p\right] \qquad \qquad
\nonumber\\
&=&\sum_{k=0}^{\infty}\frac{1}{z^{2k+1}}C_k\,\sigma^{2k} =
\frac{z}{2\pi}\int_{-2}^2 dx \frac{ \sqrt{4-x^2}}{z^2-x^2\sigma^2}=
\frac{1}{2\sigma^2}\left(z-\sqrt{z^2-4\sigma^2}\right) \qquad , \qquad
\nonumber\\
\rho(\lambda)&=&-\frac{1}{\pi} {\rm Im}\,G(\lambda+i \epsilon)=\left\{
\begin{array}{cc}\frac{1}{2 \pi \sigma^2} \sqrt{4\sigma^2 -\lambda^2}  & {\rm if} \quad 
|\lambda|<2\sigma\\
0 & {\rm if} \quad |\lambda|>2\sigma\\ 
\end{array} \right.  \nonumber\\
\label{bb.5}
\end{eqnarray}

{\bf Random graph scaling}\\
The simplest model of random graph is the ensemble $\{S_0\}$, with the probability density
 \begin{eqnarray}
s_{i,j}= \left\{ \begin{array}{cc}
1 \quad , \quad {\rm with}& {\rm probability} \quad 1/n \quad , \\
0 \quad , \quad  {\rm with}& {\rm probability} \quad 1-1/n \quad , 
\end{array}\right.
\label{bb.10}
\end{eqnarray}
Then E $[(s_{i,j})^k]=<v^k>=1/n$ for every $k$. By inserting this in the tables (\ref{b.4}),
(\ref{b.2}) and by keeping only the leading order contribution in the limit $n \to \infty$ , one finds that all the odd moments and therefore the odd powers of traces are negligible.
The limiting form of the tables become
\begin{eqnarray}
\lim_{n \to \infty} \frac{1}{n} <{\rm Tr}\, S^2>&=& \,1 \qquad , \qquad \nonumber\\
\lim_{n \to \infty} \frac{1}{n} <{\rm Tr}\, S^4>&=& \,3 \qquad , \qquad \nonumber\\
\lim_{n \to \infty} \frac{1}{n} <{\rm Tr}\, S^6>&=& \,12 \qquad , \qquad \nonumber\\
\lim_{n \to \infty} \frac{1}{n} <{\rm Tr}\, S^8>&=& \,57 \qquad , \qquad \nonumber\\
\label{bb.11}
\end{eqnarray}
These numbers reproduce the asymptotic behaviour in the limit $n \to \infty$ of the moments
E $\left[\frac{1}{n} {\rm Tr}\, S^{2p}\right]=m_p$ , evaluated by A.Khorunzhy and V.Vengerovsky \cite{ak1}. They also found a recursive relation that evaluates the asymptotic moments $m_p$ and studied their behaviour at increasing values of $p$. The same recursive relation was obtained and analyzed by M.Bauer and O.Golinelli \cite{bau}.\\

A very similar scaling is used in models of {\bf sparse matrices} :
 \begin{eqnarray}
<(s_{i,j})^k>=<v^k>=\frac{1}{n}c_k
\label{bb.12}
\end{eqnarray}
It is easy to see from the tables (\ref{b.4}), (\ref{b.2}) that with this rescaling same limit holds for matrices of the two ensembles $\{S\}$ and $\{S_0\}$) and that the odd moments vanish
$$\lim_{n \to \infty}\frac{1}{n} <{\rm Tr}S^{2p+1}>=0$$
 \begin{eqnarray}
\lim_{n \to \infty} \frac{1}{n} <{\rm Tr}\, S^2>&=& \,c_2 \qquad , \qquad   \nonumber\\
\lim_{n \to \infty} \frac{1}{n} <{\rm Tr}\, S^4>&=&c_4+2(c_2)^2  \qquad ,  \nonumber\\
\lim_{n \to \infty} \frac{1}{n} <{\rm Tr}\, S^6>&=&c_6+6c_4c_2+5(c_2)^3  \qquad ,\qquad \nonumber\\
\lim_{n \to \infty} \frac{1}{n} <{\rm Tr}\, S^8>&=&c_8+8c_6c_2+6(c_4)^2+28 c_4(c_2)^2+
14 (c_2)^4
    \qquad , \qquad \qquad
 \label{bb.13}
\end{eqnarray}

{\bf Intermediate scalings}\\
In both scalings above mentioned, the Wigner scaling and the sparse matrix scaling,
the second moment of the entries is scaled with the same power
$$<(s_{i,j})^2>=<v^2>=\frac{1}{n}c_2$$
Every moment $<v^k>$ with $k \geq 2$ , is scaled by the sparse matrix scaling 
(\ref{bb.12}) in a way
 sufficient to obtain finite asymptotic limits
$\lim_{n \to \infty} \frac{1}{n} <{\rm Tr}\, S^p> $. Furthermore the same scaling suppresses asymptotically every contribution of the first moment $<v>=c_1/n$.\\

If the first moment $<v>$ is not zero, and the matrix entries have
Wigner scaling, $<v^k>=c_k/n^{k/2}$, see eq.(\ref{cc.1}), several rescaled E$[ {\rm Tr}\, S^p]$ or
E$[ {\rm Tr}\, S_0^p]$ diverge in the $n \to \infty$ limit. I considered a simple gaussian example in the Appendix. Most authors assume the first moment $<v>=0$ together with Wigner scaling, as I did at the beginning of this section. \\

 Wigner scaling
suppresses asymptotically every contribution $<v^k>$ with $k>2$ and in this way
obtains universality.  \\

Let us parametrize possible scalings of the moments of the probability distribution of the entries 
\begin{eqnarray}
<(s_{i,j})^k>=<v^k>=\frac{1}{n^{a(k)}}c_k
\label{bb.15}
\end{eqnarray}
Fig.10 plots the rescaling power $a(p)$ on the vertical axes. The horizontal line $a(p)=1$ corresponds to the sparse matrix scaling ; the line $a(p)=p/2$, with $p \geq 2$ corresponds to Wigner scaling.\\

\noindent
\epsfig{file=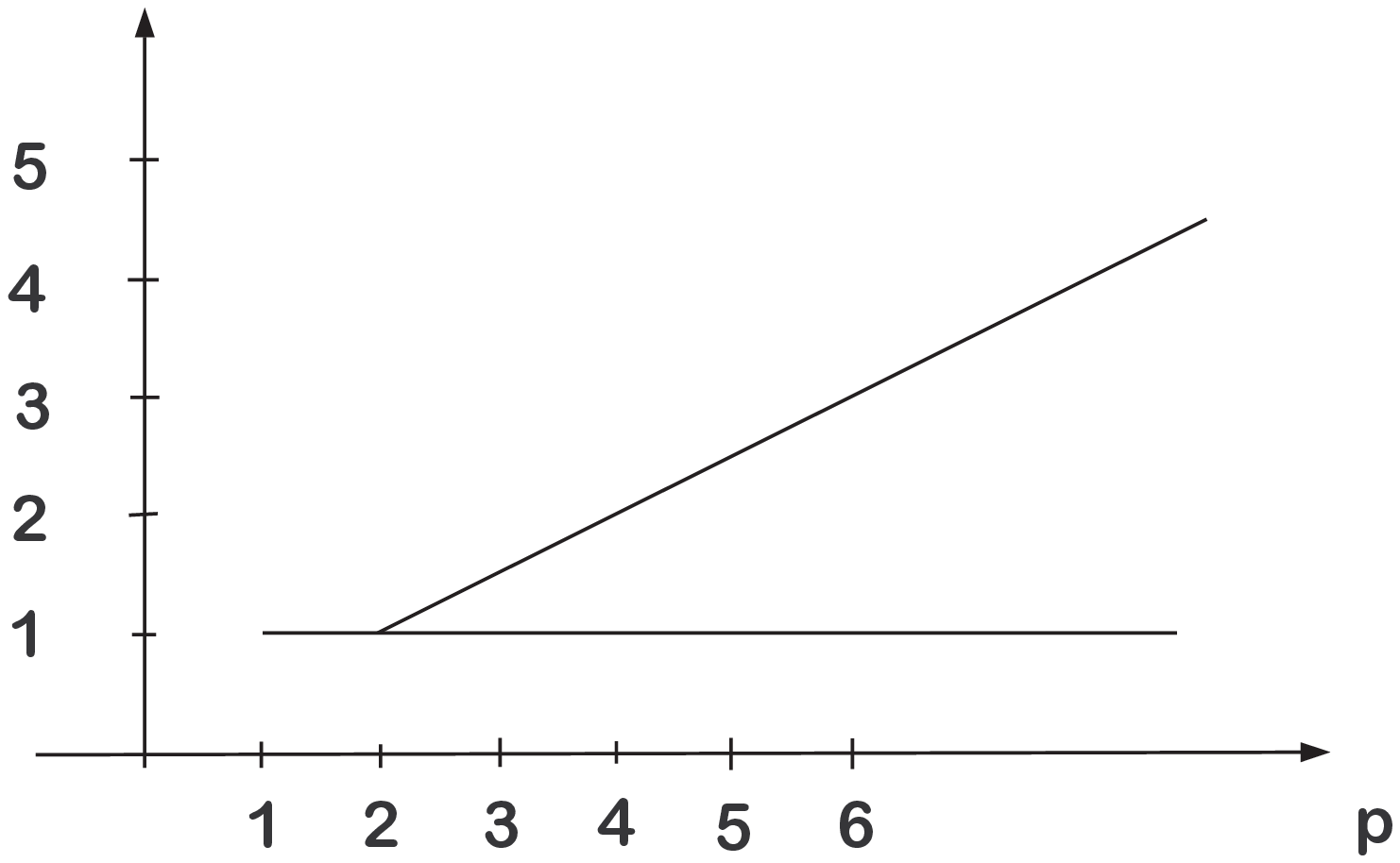,height=4.5cm}\\
 \centerline{Fig.10}\\

Any line in the sector between the two lines, that is any scaling with $a(p)=\alpha \, p+1-2\alpha$ with $0<\alpha<1/2$ and $p\geq 2$, supplemented by $<v>=0$, would obtain finite asymptotic evaluations for $ \frac{1}{n} <{\rm Tr}\, S^p> $. Our tables, up to $p=8$ suggest that {\it such asymptotic evaluations would be identical with Wigner scaling}, provided the first moment vanishes, $<v>=0$.\\

A simple example of distribution probability corresponding to intermediate scaling is
 \begin{eqnarray}
s_{i,j}= \left\{ \begin{array}{cccc}
0 \quad ,& \quad  {\rm with}& {\rm probability}& \quad 1-\frac{1}{n^{1-2\alpha}} \quad , \\
-\frac{1}{n^{\alpha}} \quad ,& \quad {\rm with}& {\rm probability}& \quad \frac{1}{2n^{1-2\alpha}} \quad , \\
\frac{1}{n^{\alpha} }\quad ,& \quad {\rm with}& {\rm probability}& \quad \frac{1}{2n^{1-2\alpha}} 
\end{array}\right.
\label{bb.18}
\end{eqnarray}
where $0<\alpha<1/2$ ; then one evaluates
 \begin{eqnarray}
< (s_{i,j})^k>= \left\{ \begin{array}{ccc}
0 \quad ,& \quad  {\rm if}& k \quad {\rm odd}\quad , \\
\frac{1}{n^{\alpha (k-2)+1}} \quad ,& \quad  {\rm if}& k \quad {\rm even}\quad 
\end{array}\right.
\label{bb.19}
\end{eqnarray}

\section{Appendix}

The effects of a non-vanishing expectation value $<v>$ may be seen in the simple model of the ensemble $\{S\}$ of  real symmetric matrices with normal probability distribution for the entries.

 The joint probability density factorizes
\begin{eqnarray}
p\left(s_{1,2} \,,\,s_{1,3}\, ,..,s_{n-1,n}\right) =\prod_{i < j}p\left(s_{i,j}\right)\qquad , \qquad  
p\left(s_{i,j}\right)=  
\frac{ 1}{\sigma \sqrt{2\pi}}e^{-(s_{ij}-v)^2/2\sigma^2}  
\qquad \qquad
 \label{rr.1}
\end{eqnarray}
 The first moments $<v^k>$ are well known
\begin{eqnarray}
&& <s_{ij}>=v \qquad , \qquad <(s_{ij})^2>= \sigma^2+v^2  \qquad , \qquad \nonumber\\
&& <(s_{ij})^3>= 3\sigma^2 v+v^3   \qquad , \qquad <(s_{ij})^4>= 3\sigma^4+6\sigma^2v^2+v^4  \qquad , \qquad \nonumber\\
&& <(s_{ij})^5>=15 \sigma^4v+10\sigma^2v^3+v^5  \qquad , \qquad <(s_{ij})^6>= 15\sigma^6+45\sigma^4v^2+15\sigma^2v^4+v^6  \qquad , \qquad \nonumber\\
&& <(s_{ij})^7>=105 \sigma^6 v+105 \sigma^4v^3+21\sigma^2v^5+v^7 \qquad , \qquad 
 \label{rr.2}
\end{eqnarray}
By inserting these moments, after Wigner rescaling $s_{ij}\to s_{ij}/\sqrt{n}$ , in the table (\ref{b.2}), one finds the large $n$ behavior of $<{\rm Tr}\, S^p>$ for $p=1, 2,..,7$.
\begin{eqnarray}
\frac{1}{n} <{\rm Tr} \,S>&=&\,v/\sqrt{n}\qquad , \qquad \nonumber\\
\frac{1}{n} <{\rm Tr} \,S^2>&=& (\sigma^2+v^2)  \qquad , \qquad \nonumber\\
\frac{1}{n} <{\rm Tr} \,S^3>&=&\,n^{1/2}\,v^3 +3n^{-1/2}\sigma^2 v\qquad , \qquad 
\nonumber\\
\frac{1}{n} <{\rm Tr} \,S^4>&=&\,n\,v^4 + 2(\sigma^2+2v^2)\sigma^2+
\frac{1}{n}(\sigma^2+4 v^2)\sigma^2+O\left(\frac{1}{n^2}\right)\qquad , \qquad 
\nonumber\\
\frac{1}{n} <{\rm Tr} \,S^5>&=&n^{3/2}v^5 + 5n^{1/2}\sigma^2 v^3+\frac{10}{n^{1/2}}(\sigma^2+v^2)\sigma^2 v+O\left(\frac{1}{n^{3/2}}\right)
\qquad ,  \nonumber\\
\frac{1}{n} <{\rm Tr} \,S^6>&=&n^2\, v^6 +  
6n \sigma^2 v^4 +(11\sigma^4+15\sigma^2v^2+18v^4)\sigma^2+O\left(\frac{1}{n}\right)
\qquad ,\nonumber\\
\frac{1}{n} <{\rm Tr} \,S^7>&=& n^{5/2}  \,v^7 + 
  7 n^{3/2}\sigma^2 \,v^5+
 7n^{1/2}(3\sigma^2+4v^2)\sigma^2 v^3 +O(1)
\qquad \qquad \qquad
\label{rr.5}
\end{eqnarray}

The large-$n$ resolvent $G(z)$ of this ensemble may be investigated with the addition law of random matrices. \cite{past1}, \cite{zee2},\cite{voi}, \cite{janik}.\\

Any real symmetric matrix of the ensemble $\{S\}$ , where $<v>$ is not equal to zero, is the sum of a random matrix of the same ensemble, with $<v>=0$ and the fixed matrix  $vJ$, where the  $J$ matrix has all entries equal to one. After Wigner scaling all the entries of the random and fixed matrices have the further scaling factor $n^{-1/2}$.\\

The resolvent $G(z)$ and its inverse function, for the fixed matrix
  $vJ n^{-1/2}$ are
\begin{eqnarray}
G(z)&=&\frac{n-1}{n}\frac{1}{z}+\frac{1}{n}\frac{1}{z-v\sqrt{n}} \qquad , \qquad \nonumber\\
z&=& \frac{1+\sqrt{n} v G + \sqrt{(1-\sqrt{n}vG)^2+4n^{-1/2}vG}}{2G}
\label{rr.7}
\end{eqnarray}

 The large$-n$ limit for the resolvent and its inverse function, for
the ensemble of gaussian matrices with $<v>=0$ correspond to the Wigner semi-circle
\begin{eqnarray}
G(z)&=&\frac{1}{2\sigma^2}\left( z-\sqrt{z^2-4\sigma^2}\right) \qquad , \nonumber\\
z&=&\frac{1}{G}+\sigma^2 G \qquad
\label{rr.6}
\end{eqnarray}
 
 According to the law of addition of random matrices, the resolvent and its inverse function, for the ensemble $\{Sn^{-1/2}\}$, for large-$n$ have the Pastur form

\begin{eqnarray}
G(z)&=&\frac{1}{n}\sum_j \frac{1}{z-\epsilon_j-\sigma^2G}=\frac{n-1}{n}\frac{1}{z-\sigma^2G}+\frac{1}{n}\frac{1}{z-v\sqrt{n}-\sigma^2 G} \qquad , \nonumber\\
 z&=&\sigma^2 G+\frac{1+\sqrt{n} v G + \sqrt{(1-\sqrt{n}vG)^2+4n^{-1/2}vG}}{2G}=\qquad \qquad
\label{rr.10}
\end{eqnarray}
On the right side of eq.(\ref{rr.10}) for the resolvent $G(z)$ , the first term dominates the second one for every $z$ finite and $n$ large, whereas the second term is dominant for $z$ very close to $v\sqrt{n}$. Then
a simple approximate form follows by considering only one term

\begin{eqnarray}
G(z)\sim \left\{
\begin{array}{ccc}
\frac{1}{2\sigma^2}\left( z-\sqrt{z^2-4(1-1/n)\sigma^2}\right) &
z \quad {\rm finite}\quad {\rm and}\quad n  \quad {\rm large}\qquad ,\\
\frac{1}{2\sigma^2}\left( z-v\sqrt{n}-\sqrt{(z-v\sqrt{n})^2-4\sigma^2/n}\right) & z\sim v\sqrt{n}+\frac{\sigma^2}{v\sqrt{n}}
\qquad 
\end{array} \right.
\nonumber
\end{eqnarray}
corresponding to the spectral densities
\begin{eqnarray}
\rho(\lambda) &\sim& \frac{1}{2\pi \sigma^2}\sqrt{4(1-1/n)\sigma^2-\lambda^2}\quad ,
\quad {\rm if} \quad |\lambda|< 2\sigma   \sqrt{(1-1/n)} \quad , \quad n \quad {\rm large}
\qquad \nonumber\\
\rho(\lambda) &\sim& \frac{1}{2\pi \sigma^2}\sqrt{4\sigma^2/n-( \lambda-v\sqrt{n})^2} \quad , \quad {\rm if}
\quad | \lambda-v\sqrt{n}|<2\sigma/ \sqrt{n}
\nonumber
\end{eqnarray}
Here the upper line reproduces the usual semi-circle spectral density, while the lower line is a very narrow peak at $\lambda=v \sqrt{n}$.\\

Finally the  eq.(\ref{rr.10}) for $z=z(G)$ is easily expanded for small values of $G$, the expansion is inverted obtaining the large $z$ expansion of the resolvent, as it follows from the addition law
\begin{eqnarray}
G(z)&=&\frac{1}{z}+\frac{v}{n^{1/2}}\frac{1}{z^2}+\frac{v^2+\sigma^2}{z^3}+
(n^{1/2}v^2+\frac{3}{n^{1/2}}\sigma^2)v\frac{1}{z^4}+\nonumber\\
&+&\left(nv^4+2(\sigma^2+2v^2)\sigma^2+\frac{2}{n}\sigma^2v^2\right)\frac{1}{z^5}+\nonumber\\
&+&\left(n^{3/2}v^5+5n^{1/2}v^3\sigma^2+5n^{-1/2}v\sigma^2(v^2+2\sigma^2)\right)\frac{1}{z^6}+\qquad \nonumber\\
&+&\left(n^2 v^6+6n \sigma^2v^4+(5\sigma^4+15\sigma^2v^2+9v^4)\sigma^2+\frac{15}{n}\sigma^4v^2\right)\frac{1}{z^7}+\qquad \nonumber\\
&+&\left(n^{7/2}v^7+7 n^{3/2}\sigma^2v^5+7n^{1/2}(2v^2+3\sigma^2)\sigma^2 v^3\right)\frac{1}{z^8}+\cdots \qquad
\label{rr.9}
\end{eqnarray}
By comparing this expansion with the direct evaluation of the large $n$ behaviour of E$({\rm Tr}\,S^p)$ given in (\ref{rr.5}), we see that the first two terms of the expansion are correctly reproduced for every $p=1,2,\cdots,7$. Analogous analysis and conclusions are abtained for the $\{S_0\}$ ensemble.\\

{\bf Acknowledgments}\\
R. De Pietri gave me precious suggestions on symbolic software and L.Molinari worked with me on much material  leading to this paper.\\
I thank the anonymous referee for critical and pertinent remarks to the first version of this paper, particularly concerning the case where $<v>$ is not equal to zero.\\

\newpage

 \large{Figure Captions.}\\

Fig.1. The left side shows the complete graph with $N$ vertices, the right side shows a $4$ steps path on it.\\

Fig.2. In both graphs of the figure, time is the horizontal axis. The left side shows the history of a path on the graph by exhibiting the visited sites. The right side is the graph of a reduced path corrisponding to the graph on the left side.\\

Fig.3. The four reduced paths which yield $<{\rm Tr}\, S_0^4>$.\\

Fig.4. The complete graph with $N$ vertices corresponding to the symmetric matrix $S$, with non-zero entries on the diagonal.\\

Fig.5. Two reduced paths of two and three steps.\\

Fig.6. One horizontal step has been added in $4$ possible ways to the $3$ steps graph of Fig.5.\\

Fig.7. The six graphs are generated by adding $2$ horizontal steps to the $2$ steps graph of Fig.5.\\

Fig.8. The left side represents the graph corresponding to the bi-diagonal matrix in eq.(2.13). The right side represents a $14$ steps paths on the graph  beginning and ending at site $j$. Time is the horizontal axis.\\

Fig.9. The one-to-one correspondence between Dyck paths and rooted trees.\\

Fig.10. The re-scaling power $a(p)$. The line $a=1$ is the re-scaling used in random graphs, the line $a(p)=p/2$ is the Wigner scaling.\\

\end{document}